\documentclass[aps,preprint,pre,array,epsfig,eqsecnum]{revtex4}
\usepackage{amsmath}
\usepackage{amsfonts}
\usepackage{natbib}
\usepackage[dvips]{graphicx}
\usepackage{epsfig} 
\usepackage[]{subfigure}
\usepackage[dvips]{color}

\def\la{\langle}
\def\ra{\rangle}
\newcommand{\beq}{\begin{equation}}
\newcommand{\eeq}{\end{equation}}

\begin{document}   
\setlength{\parindent}{0pt}

\title{A tunable solid-on-solid model of surface growth.}

\author{S.L. Narasimhan$^1$ and A. Baumgaertner$^2$  }

\affiliation{
$^1$Solid State Physics Division, Bhabha Atomic Research Centre,
Mumbai - 400085, India \\
$^2$Institute of Solid State Research, 
Research Centre J\"ulich, Germany 
}

\date{\today}

\begin{abstract} 

We have performed a detailed Monte Carlo study of a diffusionless
$(1+1)$-dimensional solid-on-solid model of particle deposition and evaporation
that not only tunes the roughness of an equilibrium surface but also
demonstrates the need for more than two exponents to characterize it. The
tunable parameter, denoted by $\mu$, in this model is the dimensionless surface
tension per unit length. For $\mu < 0$, the surface becomes increasingly spikier
and its average width grows linearly with time; for $\mu = 0$, its width grows
as $\sqrt{t}$. On the other hand, for positive $\mu$, the surface width shows
the standard scaling behavior, $\la \sigma _m(t)\ra \sim M^{\alpha}f(t/M^{\alpha
/\beta})$ where $M$ is the substrate size and $f(x) \to const \ (x^{\beta})$ for
$x$ large (small). The roughness exponent, $\alpha = 1/2$ for $\mu \leq 2$, and
$ = 3/5, 4/5 \ \& \ \sim 1$ for $\mu = 5, 6 \ \& \ 7$ respectively; the growth
exponent, $\beta = 1/4$ for $\mu \leq 2$ and $= 1/2$ for $\mu > \sim 3.5$
respectively. These exponents are different from those of the height-difference
correlation function,$\alpha ' = 1/2, \ \beta ' = 1/4$ and $z' = 2$,
for higher values of $\mu$ suggesting thereby that the surface could be self-constraining.   

\end{abstract}

\maketitle



\section{Introduction}

Stochastic evolution of rough surfaces into stationary self-affine structures has been a subject of intense study \cite{Family90,Zhang95,Krug97}. The initial growth and the asymptotic stationary value of the width, $\la \sigma _m(t)\ra$, of a surface are described by the dynamical scaling form, $\la \sigma _m(t)\ra \sim M^{\alpha}f(t/M^{\alpha /\beta})$ where $M$ is the substrate size and $f(x) \to const \ (x^{\beta})$ for $x$ large (small). The scaling exponents $\beta$ and $\alpha$ lead to a classification of the surfaces into distinct universality classes such as, for example, the Kardar-Parisi-Zhang (KPZ) class \cite{KPZ86} and the Edwards-Wilkinson (EW) class \cite{Edwards82}. 

The simplest physical process giving rise to the growth of rough surfaces is the Random Deposition (RD) of finite sized solid particles on a flat substrate; at every instant of time, the height of the surface at a randomly chosen site is incremented by the size of the particle - {\it i.e.,} a Solid-on-Solid (SOS) event occurs at a random site. Since the SOS events are statistically independent of each other, the surface growth is equivalent to a one dimensional random walk for the height variable; consequently, the surface-width, $\la \sigma _m(t)\ra$, is independent of the substrate (or system) size and increases monotonically with time, $\la \sigma _m(t)\ra \sim t^{1/2}$. 

On the other hand, if the particle sticks to the surface as soon as it makes the first contact, then the direction of growth is not always perpendicular to the substrate as it is in the case of RD process. 
Possibility of such lateral growth leads to the formation of voids in the bulk and overhangs in the growing surface. The width of the surface, however, saturates to a (substrate) size-dependent value because of a build up of spatial correlation between the surface heights. Surface grown by such a process, known as the Ballistic Deposition (BD), belongs to the KPZ class.
 
An SOS event followed by the diffusion of the deposited particle on the surface smoothens the surface. The morphology of such smoothened surfaces depends on the nature of diffusive transport taking place - namely, whether it is due to local height differences \cite{Edwards82} or due to local curvature \cite{LaiDas}. Most of these models use simple phenomenological or {\it adhoc} rules for the diffusion of deposited particles on the surface. The asymptotically saturated surface is referred to as the {\it equilibrium} surface, not in the thermodynamic sense but in the sense that its width has attained a stationary value.   

The growth of such a surface can also be described by the height-difference correlation function (HDCF), $G_M(k,t)$, that is defined in terms of the average differences in the heights of sites separated by a distance $k$ \cite{KK,KKA} - namely, $G_M(k,t) \equiv \la [h(i+k,t)-h(i,t)]^2\ra $ where the averaging is over all reference sites $i$ and also over independent realizations of the surface. It has been shown \cite{KK,KKA} that $G_M(k,t) \sim k^{2\alpha '}f_G(k/\xi _G(t))$, where $\alpha '$ is known as the {\it wandering} exponent; initially ($t \ll M^{z'}$), the correlation length, $\xi _G(t)$, increases with time as $\xi _G(t) \sim t^{1/z'}$ but saturates to a value proportional to $M$ for $t \gg M^{z'}$. Normally, for SOS models with local growth rules, $\alpha ' = \alpha$ and $z' = z$. However, for SOS models with global constraints on surface heights such as, for example, the Suppressed Restricted Curvature model \cite{JK}, $\alpha ' \neq \alpha$ and $z' \neq z$, eventhough their ratios may be equal ($\alpha '/z' = \beta = \alpha/z$). An interesting question to ask is whether there is a growth process that necessitates the need for more than two exponents to characterize an equilibrium surface.

In this paper, we present a detailed Monte Carlo study of a diffusionless $(1+1)$-dimensional model of particle deposition and evaporation that not only generates an equilibrium surface with tunable roughness but also demonstrates the need for more than two exponents to characterize it. This model is based on the observation that adding or removing a particle is equivalent to changing the area of the surface by one basic unit. It has a single tunable parameter, denoted by $\mu$, the magnitude of which measures the strength, in units of thermal energy, of this change. 

We show that the morphology of the one dimensional surface changes drastically from spiky to rough as $\mu$, the surface tension parameter, changes sign. More interestingly, we show that different ranges of $\mu$ values correspond to different universality classes - (i) EW class for $\mu$ in the range $0 < \mu \leq 2$; (ii) growth exponent $\beta = 1/2$ for $\mu \geq 3.5$ while the roughness exponent $\alpha$ increases monotonically with $\mu$ - namely, $\alpha \sim 1/2, 3/5, 4/5$ and $1$ for $\mu = 3.5, 5, 6$ and $7$ respectively. The Monte Carlo data presented here suggest that the parameter $\mu$ can tune the exponents $\alpha$ and $\beta$. On the other hand, the exponents $\alpha '$, $z'$ and $\beta ' ( = \alpha '/z')$ of the HDCF, $G_M(k,t)$, retain their EW values independent of $\mu$. They are, thus, different from those of the surface width eventhough no arbitrary global constraint is imposed on surface heights. This is in contrast to the Suppressed Restricted Curvature model \cite{JK} that explicitly invokes a global constraint on the extremal height of the surface. The paper is organized as follows. 

In section II, we introduce and discuss the model in detail; in section III, we present a detailed Monte Carlo study of the one dimensional model and finally, in section IV, summarize the results.

\section{A $(1+1)$-dimensional SOS deposition-evaporation model.}

Let ${\cal M}(t)$ denote a surface configuration at time $t$ growing on a one dimensional lattice substrate consisting of $M$ sites. The particle being deposited (or removed) is assumed to be a unit square so that the surface ${\cal M}(t)$ is actually a collection of integer-valued variables ({\it heights}), $\{h_j(t)\mid j = 1, 2,\cdots ,M\}$. The contour length of ${\cal M}(t)$ can then be written as,
\begin{equation}
L_{{\cal M}}(t) = M + \sum _{i=1}^M \mid h_{i+1}(t) - h_i(t)\mid
\label{eq:L}
\end{equation} 
with periodic boundary codition, $h_{M+1}(t) = h_1(t)$, invoked.

During the growth process, if a particle is deposited or removed at a randomly chosen site $j$, then the change in the contour length because of a change in $h_j(t)$ by $\pm 1$ is given by
\begin{eqnarray}
\Delta L_{\cal M}(t) & = & L_{\cal M}'(t) - L_{\cal M}(t) \\  
            & = & \mid h_{j+1}(t) - h'_j(t)\mid + \mid h'_j(t) - h_{j-1}(t)\mid   \nonumber \\
            &   & - \mid h_{j+1}(t) - h_j(t)\mid - \mid h_j(t) + h_{j-1}(t)\mid
\label{eq:dL}
\end{eqnarray}
where $h'_j(t) \equiv h_j(t-1) \pm 1$, say with probability $b$ or $1-b$ respectively; periodic boundary condition may be imposed for the edge sites $1$ and $M$. It is clear from this definition that $\Delta L(t) \in \{ 2, 0, -2\}$. 

Acceptance of the new value, $h'_j$, of height at site $j$ could be on the basis of the standard {\it Metropolis} criterion - namely, that it is accepted with probability $p[h_j] = \mbox{min}(1, e^{-\mu \Delta L(t)})$, where $\mu$ is a tunable parameter in units of $k_BT$. The surface evolves stochastically as the sites are scanned and a choice for random deposition or removal of a particle is exercised.

The excess contour length, $L_{{\cal M}}(t) - M$, defined in Eq.(1) may be recognized as the energy of a solid-on-solid (SOS) model \cite{Nelson04}, 
\beq
E = \mu \sum _{<ij>} \mid h_i(t) - h_j(t) \mid = \mu (L_{{\cal M}}(t) - M)
\label{eq:energy}
\eeq
with which the surface configuration ${\cal M}$ is weighted. It is clear that $\mu$ may be interpreted as a microscopic surface energy (``surface tension''), and the summation is over nearest neighbor pairs.
In fact, it was used \cite{Temperley,Hilhorst} for studying the influence of an external field on the equilibrium fluctuations of an overhangs-free interface in the Ising model. This $(1+1)$-dimensional model may also be considered as that of a directed polymer with fluctuating number of monomers. 

\begin{figure}
\includegraphics[width=4.0in,height=3.0in]{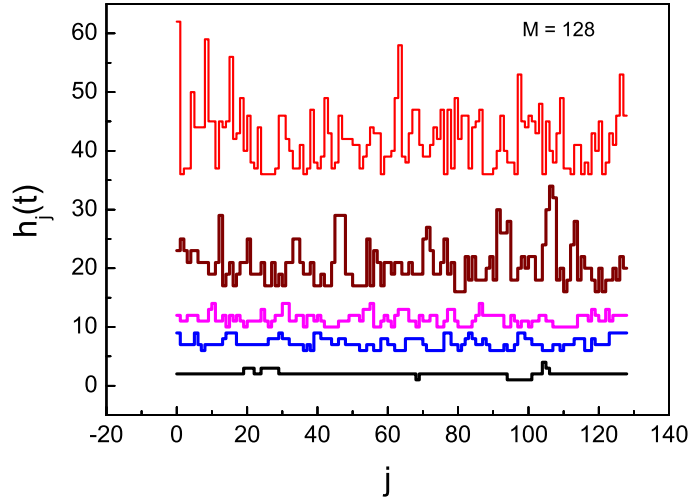}
\vspace{-0.4in}
\caption{Surface profiles after $20000$ Monte Carlo sweeps for $\mu = -0.1, 0.0, 0.5, 1.0$ and $2.0$ (from top to bottom). Substrate size $M = 128$}
\label{fig:SurfProf} 
\end{figure}

   
Typical surface profiles obtained after $20000$ Monte Carlo sweeps are shown in Fig.(\ref{fig:SurfProf}) for $\mu$ in the range $-0.1$ to $2.0$. They are smoother for positive than for negative $\mu$. In fact, the surface becomes spikier as $\mu$ becomes more negative. In all these cases, the mean height of the surface, $\la h_m \ra$, does not change because deposition and evaporation are equally likely events.
%
%

For a given realization of the surface, the width (or equivalently, the roughness) is given by the mean square fluctuation in the surface heights,
\begin{equation}
\sigma _{\cal M}^2(t) = \frac{1}{M}\sum _{j=1}^M [h_j(t) - {\bar h}(t)]^2
\label{eq:sigma}
\end{equation}
where ${\bar h}(t)$ denotes the center of mass, or equivalently, the mean height of the surface:
\begin{equation}
{\bar h}(t) = \frac{1}{M}\sum _{j=1}^M h_j(t)
\label{eq:hbar}
\end{equation}
Averages over a number of realizations may be denoted by $\la \sigma _M^2(t)\ra$ and $\la h_M(t)\rangle$ respectively.

Since the model is defined in terms of the contour length, $L$, temporal behavior of the fluctuations in $L$ should be of interest. However, it is not difficult to see that they are directly related to those of surface heights:

Let us denote the deviation of a column height from the mean height (center of mass) of the surface by $H_j(t) = h_j(t) - {\bar h}(t)$. Then we have, from Eq.(\ref{eq:L}),
\begin{eqnarray}
\frac{1}{2}\la (\delta L(t))^2 \ra 
           & \equiv & \la (L(t)-M)^2\ra \\
           & = & \sum _{i=1}^M \la (H_{i+1}(t)-H_i(t))^2\ra + X(t) \\
            & = & 2M \la \sigma _M^2(t)\ra [1 - C_1(t)] + X(t)
\label{eq:deltaL}
\end{eqnarray}
where 
\begin{equation}
X(t) \equiv 2\left\la \sum _{i=2}^M \mid H_i(t)-H_{i-1}(t)\mid \left( 
                 \sum _{j=1}^{M-i+1}\mid H_{j+1}(t)-H_j(t)\mid \right) \right\ra
\label{eq:X1}
\end{equation}
consistes of terms proportional to the correlation functions, $C_1(t), C_2(t),\cdots$, and hence can be written as
\begin{equation}
X(t) = 2M \la \sigma _M^2(t) \ra \ f[C_1(t),C_2(t),C_3(t)\cdots ]
\label{eq:X2}
\end{equation}
The correlation functions, $C_1(t), C_2(t),\cdots$, are given by the standard definition,
\begin{equation}
C_k(t) = \frac{\la \sum _{j=1}^M H_{j+k}(t)H_j(t) \ra }{M\la \sigma _M^2(t)\ra};\quad k = 1, 2,\cdots
\label{eq:corC}
\end{equation}
In the case of a pure random deposition-evaporation process, the individual columns are statistically uncorrelated with each other and so, we have the simple result,
\begin{equation}
\frac{1}{2}\la (\delta L(t))^2 \ra \propto \la \sigma _M^2(t)\ra \propto t
\label{eq:delL}
\end{equation}
In all the other cases ($\mu \neq 0$), the time-development of the correlation functions cannot be ignored. 

\section{Results and Discussions}

Starting from the initial straight line configuration, we have monitored the build-up of the surface roughness by estimating the time-dependence of the root mean squared fluctuation, $\la \sigma _m(t)\ra = \la \sigma _M^2(t)\ra^{1/2}$, averaged over many independent runs employing periodic boundary conditions.

\begin{figure} [hbt]
  \begin{center}
    \subfigure[]{\includegraphics [width=0.48\textwidth] 
    {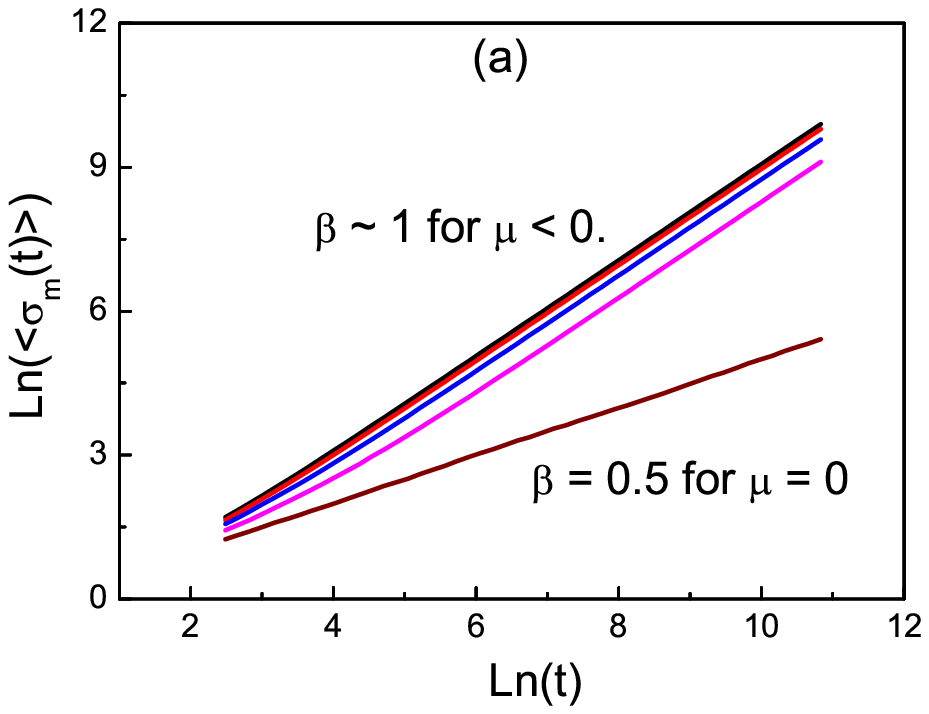}}
    \subfigure[]{\includegraphics [width=0.48\textwidth] 
    {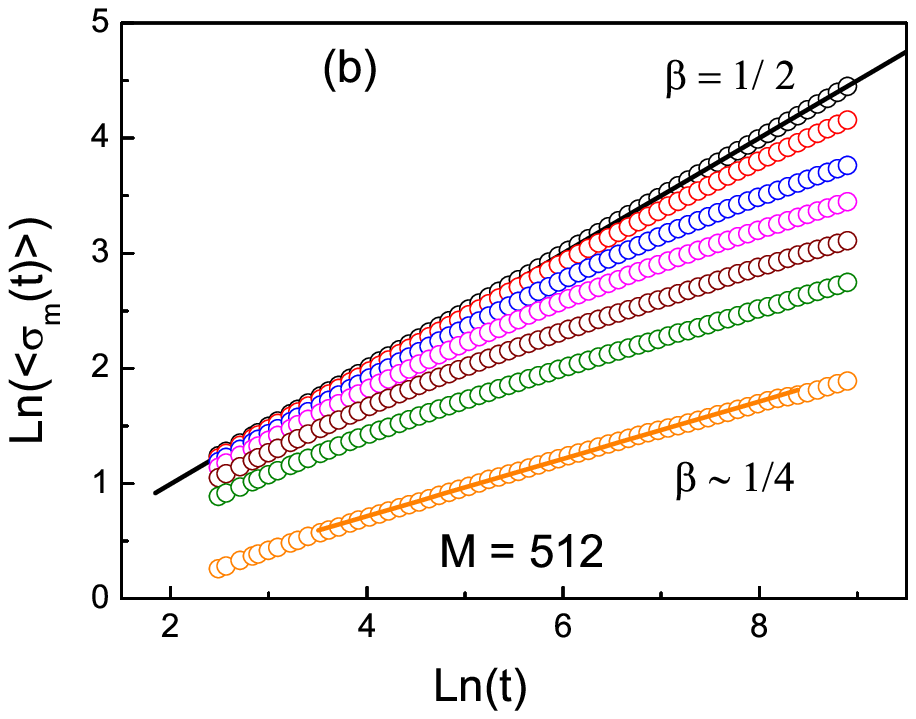}}    
\caption{ (a) Time-development of root mean-square fluctuation, $\sigma _m(t)$,
 for $\mu = -1.0, -0.75, -0.5, -0.25$ and $0.0$ (top to bottom). 
 System size $M = 128$. Slope of the line, $\beta = 1/2$ for $\mu = 0$, 
 and is asymptotically equal to unity for the other negative values of $\mu$. 
(b) $\sigma _m(t)$ for positive values of 
$\mu = 0.0, 0.1, 0.03125, 0.0625, 0.125, 0.25$ and $0.5$ (top to bottom). 
System size $M = 512$. Clearly, $\beta = 1/4$ for $\mu > 0$. 
}
\label{fig:sigmat1}
  \end{center}
\end{figure}

We have presented in Fig.(\ref{fig:sigmat1}) the Monte Carlo estimates of $\la \sigma _m(t)\ra$ for both positive and negative values of $\mu$. It is clear that $\la \sigma _m(t)\ra$ is proportional to $t$ for $\mu < 0$, whereas it is proportional to $t^{1/4}$ for $\mu > 0$. This suggests that the surface may belong to the EW class for positive $\mu$, in which case $\la \sigma _m(t)\ra$ will have to saturate asymptotically to a value that scales with the system size $M$.  

Following previous works \cite{Family86,Family90},
the finite size scaling of $\la \sigma _m(t) \ra$ for  $\mu >0$ is given by
\beq
\la \sigma _m(t) \ra = M^{\alpha}~f(t M^{-z})~ \sim \left\{
                                                    \begin{array}{ll}
                                                          t^{\beta} & \mbox{: } t \ll M^z\\
                                                          M^{\alpha} \qquad & \mbox{: } t \gg M^z.
                                                    \end{array}
                                                    \right.
                                             ;~~~~ \beta = \alpha / z,
\label{eq:sigmaScal}
\eeq
where $f(x) \sim x = const$ for $x \gg 1$ and $f(x) \sim x^{\alpha / z}$ for $x \ll 1$.

\vspace*{0.5cm}
\subsection{Dynamics of roughening ($0 < \mu \leq 2$)}

\begin{figure} [hbt] 
  \begin{center}
    \subfigure[]{\includegraphics [width=0.48\textwidth] 
    {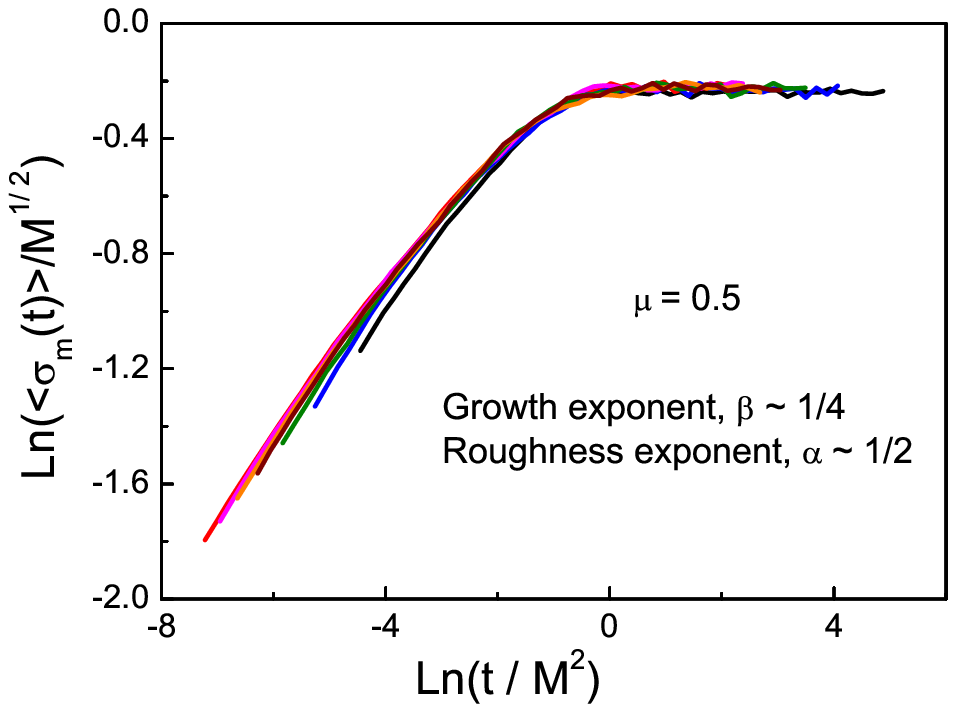}}
    \subfigure[]{\includegraphics [width=0.48\textwidth] 
    {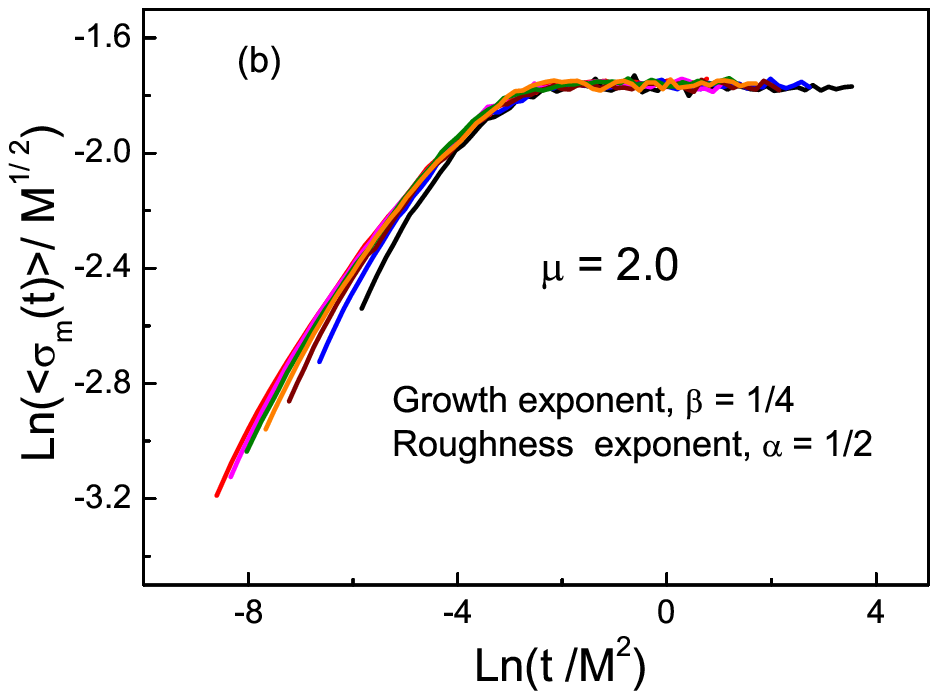}}    
\caption{Data Collapse for root mean squared height-fluctuation. 
(a) $\mu = 0.5$ and system sizes $M = 32$ to $128$ in steps of $16$. 
(b) $\mu = 2$ and system sizes $M = 64$ to $256$ in steps of $32$. 
The exponents are $\alpha = 1/2$ and $\beta = 1/4$ in either case.}
\label{fig:sigmaScalEW}
  \end{center}
\end{figure}

The scaling analysis of our simulation results, presented in the form of data 
collapse for $\mu = 0.5$ and $2.0$ in Fig.(\ref{fig:sigmaScalEW}), clearly 
indicate that the exponents have the values expected of the 
Edwards-Wilkinson (EW) class \cite{Edwards82,Family86,KPZ86} 
for $0< \mu \leq 2$ - namely, 
$\beta = 1/4$, $\alpha = 1/2$, and $z = 2$.
%
%
%
%

\begin{figure} [hbt] 
  \begin{center}

    \subfigure[]{\includegraphics [width=0.48\textwidth] 
    {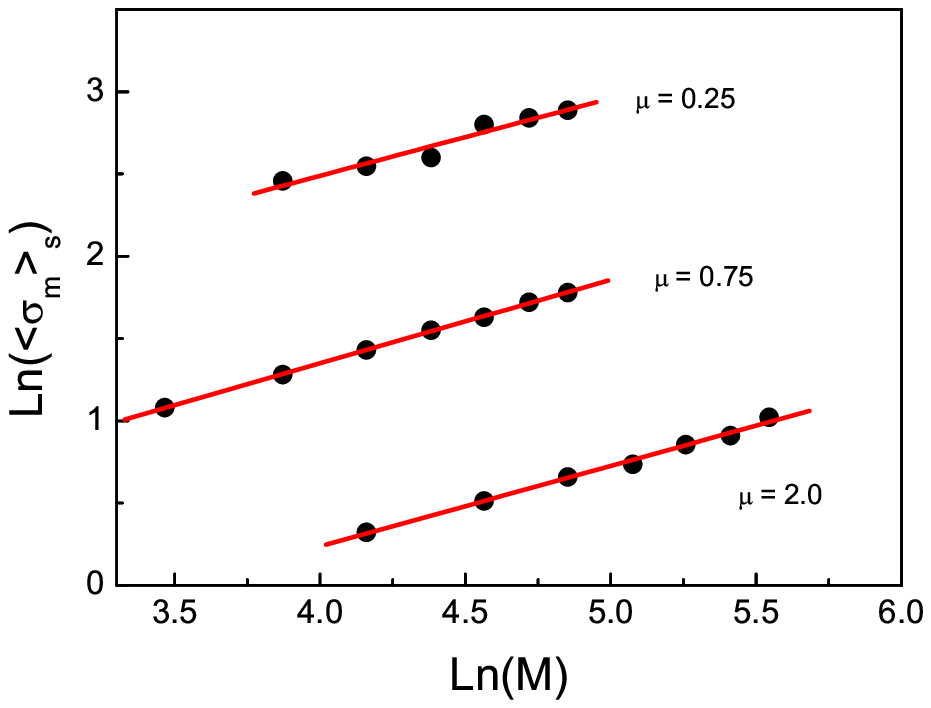}}
    \subfigure[]{\includegraphics [width=0.48\textwidth] 
    {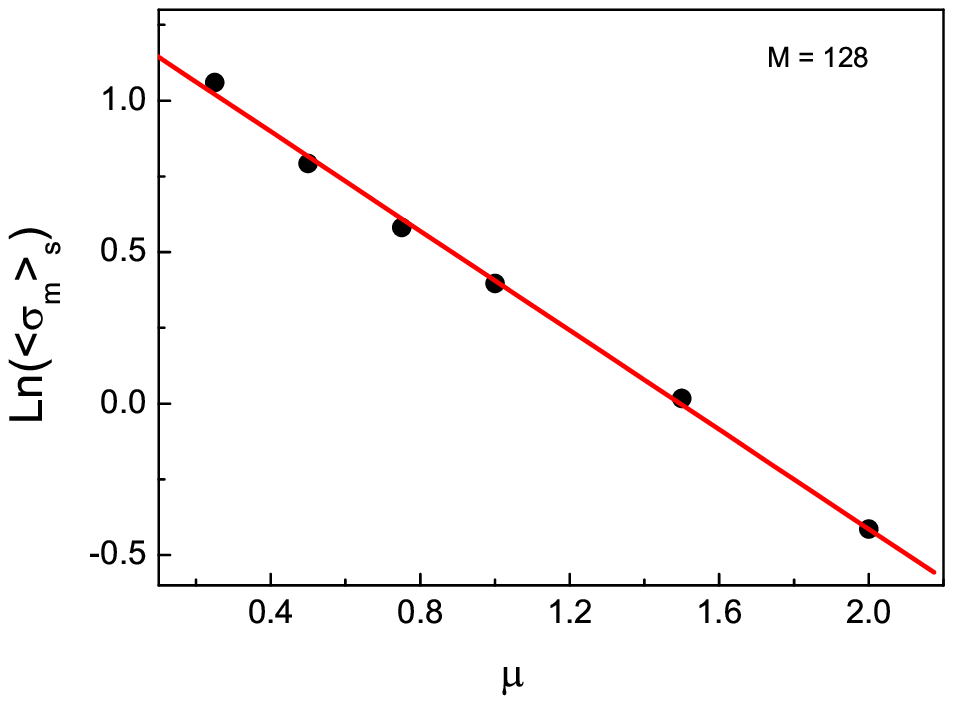}}    
    \caption{(a) Average saturation values, $\la \sigma _m \ra_s$ as
    functions of sytstem size $M$ for $\mu$ = 0.25, 0.75 and 2.0. Slope of
    the fitted straight lines is $1/2$ within errors. (b)
    Dependence of the equilibrium values $\la \sigma _m \ra_s$ on $\mu$
    for a system size $M = 128$.
    } 
    \label{fig:sigmaM}
  \end{center}
\end{figure}

The $\mu$-dependence of the saturated value, $\la \sigma _m \ra_s$, is reflected in the proportionality factor, $B(\mu)$:
\beq
\la \sigma _m \ra_s = B(\mu)M^{\alpha}
\eeq
where $\alpha = 1/2$ for $0 < \mu \leq 2$. Numerical evidence for this is presented in Fig.(\ref{fig:sigmaM}a) in which size dependence of $\la \sigma _m \ra_s$ is shown for various values of $\mu$. Moreover, data presented in Fig.(\ref{fig:sigmaM}b) for system size $M = 128$ suggest that $B(\mu)$ is an exponentially decaying function, 
\beq
B(\mu) = B_0~e^{-\mu / \mu_B},
\eeq
where the constants are $B_0 = e^{1.23 \pm 0.02}$ and $\mu_B = 1.22 \pm 0.03$.

\begin{figure} [hbt] 
  \begin{center}       
    \subfigure[]{\includegraphics [width=0.48\textwidth] 
    {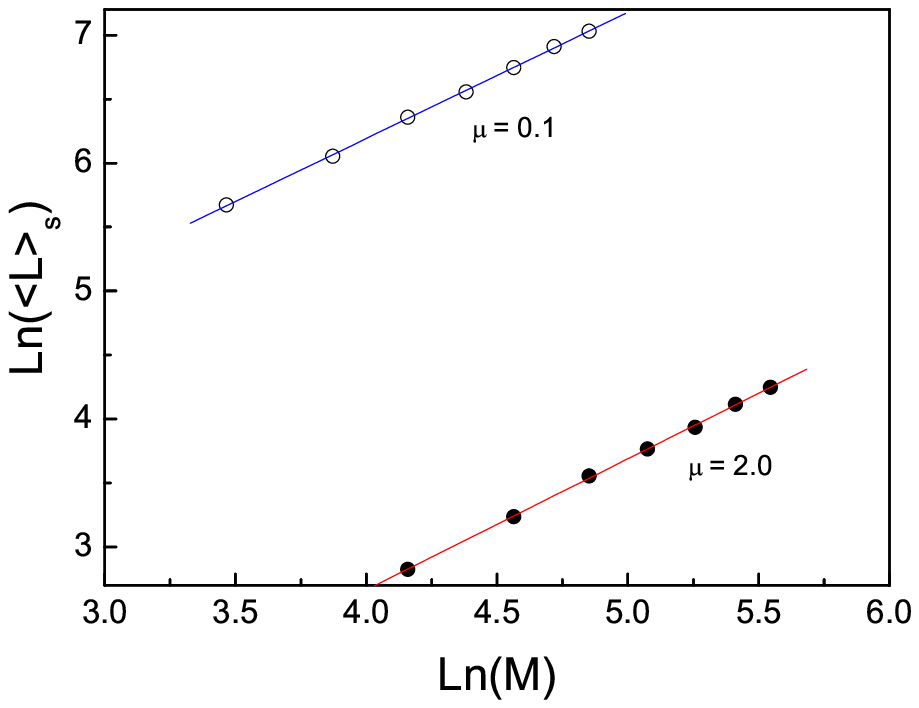}}
    \subfigure[]{\includegraphics [width=0.48\textwidth] 
    {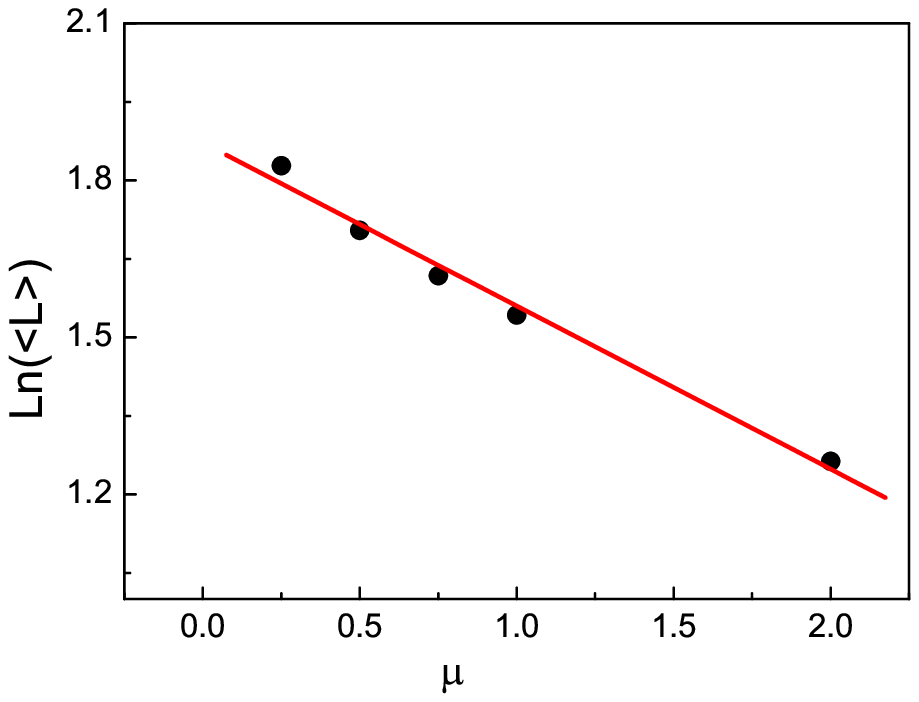}}             

    \caption{
    (a) Size-dependence of the saturation values of the average
    contour length $\la L \ra _s$ 
    for $\mu=0.1$ and $\mu=2.0$. In either case $\la L \ra \propto M$; the $\mu$-dependence is only in the proportionality constant.
    (b) Semi-log plot of the average contour length $\la L \ra _s$ for $M = 128$ fixed as function of surface tension $\mu$.  
    } 
    \label{fig:contour}
  \end{center}
\end{figure}

Since fluctuation of contour length is linearly related to that of surface heights (Eq.(\ref{eq:deltaL})), it will be of interest to know the dependence of the average contour length, $\la L \ra$, on system size and $\mu$.
Data presented in Fig.(\ref{fig:contour}a) for $\mu = 0.1$ and $2$ suggest that $\la L \ra$ is proportional to the system size:  
\beq
\la L \ra = A(\mu) M.
\eeq
The prefactor, $A(\mu)$, is an exponentially decaying function of $\mu$, as is evident from
Fig.(\ref{fig:contour}b) for system size $M = 128$:
\beq
A(\mu) = A_0~ e^{-\mu / \mu_A},
\eeq
where the constants are $A_0 = 6.49 \pm 0.13$ and $\mu_A = 3.2 \pm 0.2$.

\vspace*{0.5cm}
\subsection{Dynamics of roughening ($\mu > 2$)}

Interestingly, we find that the exponents, $\alpha$ and $\beta$, do not retain 
their EW values ($1/2$ and $1/4$ respectively) as $\mu$ is increased beyond 
the value $2$. This may not be surprising for the following reason. 

Since the probability of growth, and hence of increasing the contour length, at any given site is proportional to $e^{-2\mu}$ (Section II), larger the value of $\mu$ lesser will be the probability for growth. In the extreme limit $\mu \to \infty$, the surface hardly evolves from its intial flat morphology and at best may acquire a few steps in the long time limit $t \to \infty$. In other words, it will remain mostly flat, save for a few steps here and there, and so will not resemble an EW surface. It is therefore reasonable to expect that $\alpha$ will change from its EW value to that corresponding to a surface with flat morphology as we increase $\mu$.

\begin{figure} [hbt] 
  \begin{center}

    \subfigure[]{\includegraphics [width=0.48\textwidth] 
    {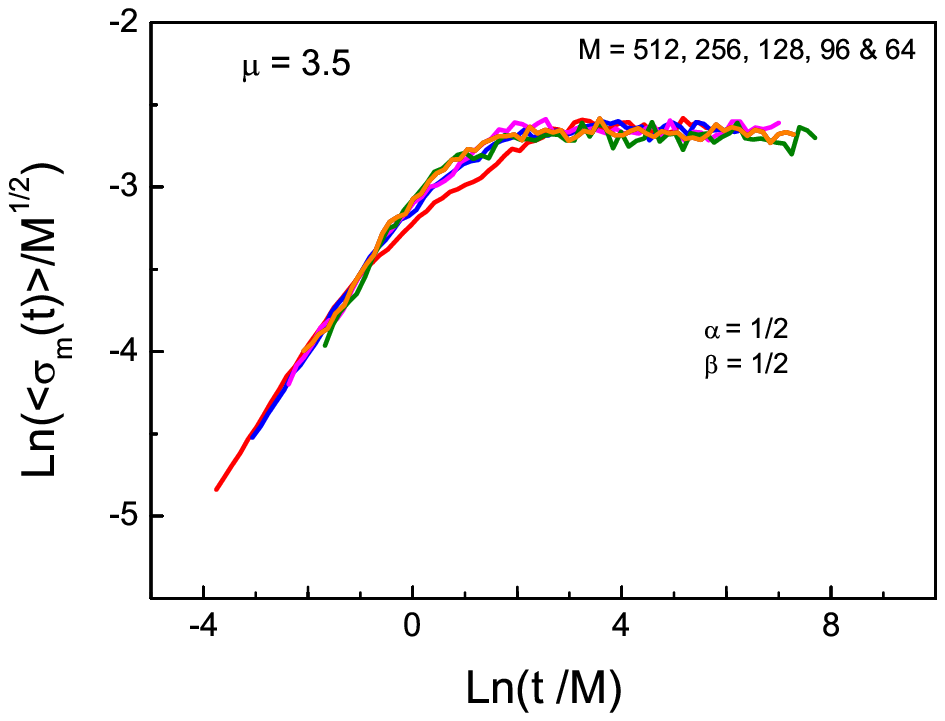}}
    \subfigure[]{\includegraphics [width=0.48\textwidth] 
    {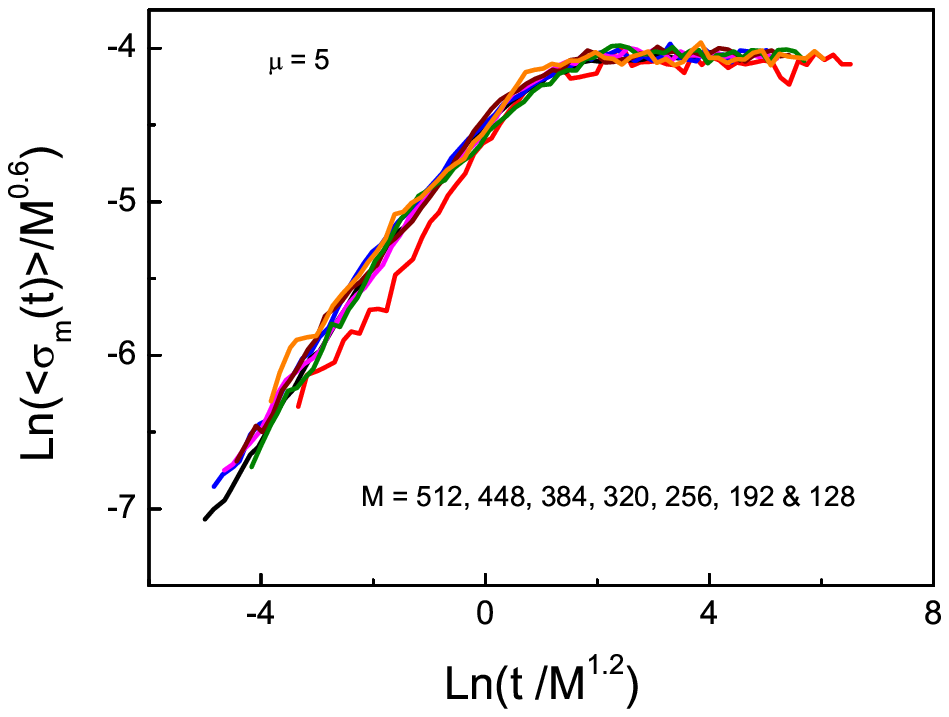}} 
    \subfigure[]{\includegraphics [width=0.48\textwidth] 
    {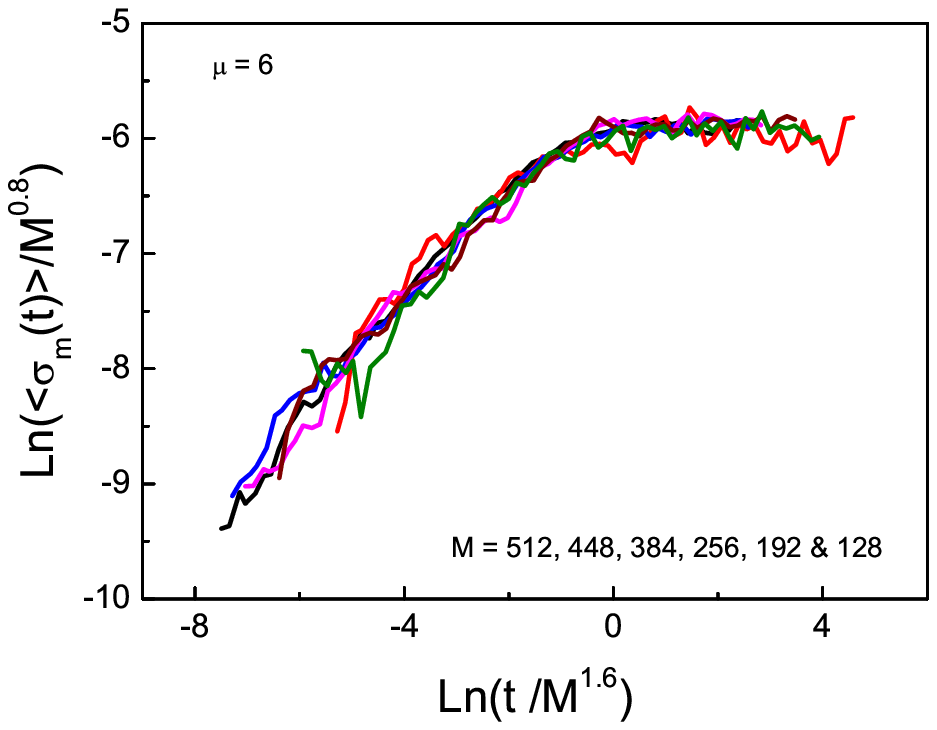}}
    \subfigure[]{\includegraphics [width=0.48\textwidth] 
    {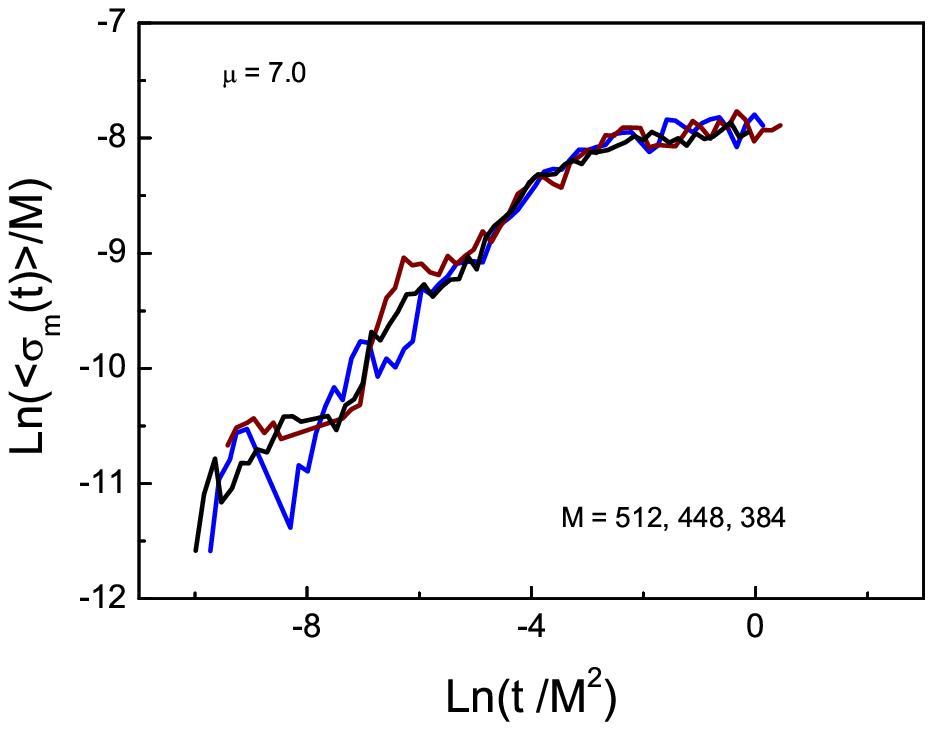}}       
    \caption{    Data collapse for various values of $M$ ( $ = 512, 448, 384, 320, 256, 192$ and $128$)   
                 (a) $\mu = 3.5$, and the exponent values are $\beta = 1/2$ and $\alpha = 1/2$. 
                 (b) $\mu = 5$, and the exponent values are $\beta = 1/2$ and $\alpha \sim 0.6$.
                 (c) $\mu = 6$, and the exponent values are $\beta = 1/2$ and $\alpha \sim 0.8$.
                 (d) $\mu = 7$; data presented for $M = 512, 448$ and $384$ $\mu = 5$. The exponent values 
                     are $\beta = 1/2$ and $\alpha \sim 1$; 
    } 
    \label{fig:SigmaScaleEW1}
  \end{center}
\end{figure}

In Fig.(\ref{fig:SigmaScaleEW1}), we have presented data collapse for $\mu = 3.5, 5, 6$ and $7$. We see that the collapse obtained is reasonably good for system sizes ranging from $M = 128$ to $512$, though the statistics becomes poorer especially for small system sizes at higher values of $\mu$; because of small $e^{-2\mu}$, larger systems and longer time spans are needed for acceptable statistics on the build-up of fluctuations.  That is why, collapse is shown Fig.(\ref{fig:SigmaScaleEW1}d) for only three values of $M$ ( = $512, 448$ and $384$) at $\mu = 7$.

The growth exponent $\beta = 1/2$ for all these values of $\mu$, whereas the roughness exponent $\alpha$ increases monotonically with $\mu$ ($\alpha \sim 1/2, 0.6, 0.8$ and $1$ for $\mu = 3.5, 5, 6$ and $7$ respectively). Larger the value of $\mu$, lesser will be the probability for a change in the surface height that leads to an increase in the contour length ({\it i.e.,} surface area in one dimension); consequently, we expect the growth of height fluctuations to be slow. Yet, $\beta$ has increased from its EW value ($ = 1/4$ for $\mu \leq 2$) to that corresponding to Random Deposition ($= 1/2$ for $\mu \geq 3.5$). In this context, it will be of interest to look at the height-height correlation data as well.

\vspace*{0.5cm}
\subsection{Height-height correlations.}

The height-difference correlation function (HDCF) \cite{KK,KKA} of a surface configuration at time $t$ - defined by the set of heights, $\{h_i(t)\mid i = 1,2,\cdots ,M\}$ - is given by
\begin{equation}
G_M(k,t) = \frac{1}{M}\sum _{j=1}^M \la(h_{j+k}(t) - h_j(t))^2\ra
\label{eq:corG}
\end{equation}
It shows a scaling behavior,
\begin{equation}
G_M(k,t) \sim k^{2\alpha '}f_G\left(\frac{k}{\xi _G(t)}\right); \quad k = 1, 2, \cdots M/2
\label{eq:scaleG}
\end{equation}
with respect to the lag variable $k$. The exponent $\alpha '$ is known as the "wandering" exponent. Normally, after growing as $t^{1/z'}$ at early times, the correlation length $\xi _G(t)$ saturates to a value proportional to the system size:
\begin{equation}
\xi _G(M,t) \sim \left\{
                        \begin{array}{ll}
                               t^{1/z'} & \mbox{ for } t^{1/z'} \ll M\\
                               M & \mbox{ for } t^{1/z'} \gg M.
                        \end{array}
                 \right.
\label{eq:CorLenG}
\end{equation}
For SOS models with local growth rules, $\alpha ' = \alpha$ and $z' = z$.

It is clear from Eqs.(\ref{eq:corG}, \ref{eq:sigma}) that $G_M(k,t)$ and $\la \sigma _M^2(t)\ra$ are related to each other by the following exact identity:
\begin{eqnarray}
G_M(k,t) & = & \frac{1}{M}\sum _{j=1}^M \la(h_{j+k}(t) - h_j(t))^2\ra \nonumber \\
       & = & \frac{1}{M}\sum _{j=1}^M \la({\hat h}_{j+k}(t) - {\hat h}_j(t))^2\ra \nonumber \\
       & = & 2(\la \sigma _M^2(t)\ra - C_M(k,t));\quad k = 1, 2, \cdots , M
\label{eq:corGsigma}
\end{eqnarray}
where ${\hat h}_i(t) \equiv h_i(t) - {\bar h}_i(t)$ and $C_M(k,t)$ is the height-fluctuation correlation function (HFCF), given by the standard definition, 
\begin{equation}
C_M(k,t) = \frac{1}{M}\sum _{j=0}^M \la{\hat h}_{j+k}(t){\hat h}_j(t)\ra ; 
                                    \quad C_M(0,t) = \la \sigma _M^2(t)\ra
\label{eq:corC_M}
\end{equation}
From Eq.(\ref{eq:corG}), it follows that 
\begin{eqnarray}
G_M(0,t) & = & 0 = G_M(M,t) \\
G_M(M/2-k,t) & = & G_M(M/2+k,t); \quad k = 1, 2, \cdots , M/2-1
\label{eq:Gsymm}
\end{eqnarray}
which implies that $C_M(k,t)$ is also symmetric about $M/2$, namely, 
\begin{equation}
C_M(M/2-k,t) = C_M(M/2+k,t)
\label{eq:Csymm}
\end{equation}
Moreover, summing $G_M(k,t)$ over $k$, we get
\begin{equation}
\frac{1}{M}\sum _{k=1}^M G_M(k,t) =  2\la \sigma _M^2(t)\ra
\label{eq:Gsum}
\end{equation}
which, together with the identity Eq.(\ref{eq:corGsigma}), implies that
\begin{equation}
\frac{1}{M}\sum _{k=1}^M C_M(k,t) = 0
\label{eq:Csum}
\end{equation}

Monte Carlo averages of $C_M(k,t)$, obtained after a time $t = e^{12.5}$ and normalized with respect to $C_M(0,t) \equiv \la \sigma _M^2(t)\ra$, are presented in Fig.(\ref{fig:corC512}) for a system of size $M = 512$ and for $\mu = 1/2 \ \& \ 5$. It is quite evident that $C_M(k,t)$, is an exponentially decaying function over a reasonable range of $k$:
\begin{equation}
C_M(k,t) = A_C e^{-k/\xi _C(M,t)}
\label{eq:corCexp}
\end{equation}
where the proportionality constant $A_C$ will depend on $\mu$ while the correlation length, $\xi _C(M,t)$, will depend on $M$ and $t$. Monte Carlo estimates of the correlation length, $\xi _C(M,t)$, are presented, and compared with those of $\xi _G(M,t)$ in Fig.(\ref{fig:CorLenXiGC}a,b) for $\mu = 1/2 \ \& \ 5$. It is clear that
\begin{equation}
\xi _C(M,t) \propto \xi _G(M,t) \sim \left\{
                        \begin{array}{ll}
                               t^{1/2} & \mbox{ for } t^{1/2} \ll M\\
                               M & \mbox{ for } t^{1/2} \gg M.
                        \end{array}
                 \right.
\label{eq:CorLenC}
\end{equation}
In view of Eq.({\ref{eq:CorLenG}), this implies that $z' = 2$. 

The exponential decay of the HFCF, $C_M(k,t)$, from a positive value $\la \sigma _M^2(t)\ra$ at $k=0$, taken with the equality Eq.(\ref{eq:Csum}), implies that $C_M(k,t)$ has to be negative over a certain range $k \in [M/2-k_0, M/2+k_0]$ for some $k_0$. In turn, this implies (by Eq.(\ref{eq:corGsigma})) that $G_M(k,t)$ has to be an increasing function of $k$. In fact, $G_M(k,t)$ has been shown to have the following scaling behavior \cite{KK,KKA}:

For $t \gg M^z$,
\begin{equation}
G_M(k,t) \sim \left\{
                    \begin{array}{ll}
                          k^{2\alpha '} & \mbox{ for } k \ll M\\
                          M^{2\alpha '} & \mbox{ for } k \approx M/2.
                    \end{array}
             \right.
\label{eq:Gscale1}
\end{equation}
and for $t \ll M^z$,
\begin{equation}
G_M(k,t) \sim \left\{
                    \begin{array}{ll}
                          k^{2\alpha '} & \mbox{ for } k \ll t^{1/z'}\\
                          t^{2\beta '} & \mbox{ for } k \gg t^{1/z'}.
                    \end{array}
             \right.
\label{eq:Gscale2}
\end{equation}

Monte Carlo estimates of $G_M(k,t)$ at various times for a system of size $M = 512$ and for $\mu = 1/2 \ \& \ 5$ are presented in Fig.(\ref{fig:corG512collapse}a) in the form of a collapse diagram suggested by the above scaling forms. The corresponding saturation values are shown in Fig.(\ref{fig:corG512collapse}b). Within numerical errors, we have $\alpha ' = 1/2, \ \beta ' = 1/4$ and $z' = 2$. It must be noted that these values are independent of $\mu$, in striking contrast to those of the surface width. For example, for $\mu = 5$, $\alpha \approx 0.6, \ \beta = 1/2$ and $z \approx 1.2$. 

In order to show that this is not a numerical artifact, we consider Eq.(\ref{eq:corGsigma}) in the initial time regime $t \ll M^z$. For small lag distances (say $k \ll M/4$), the HFCF $C_M(k,t)$ is an exponentially decaying function, Eq.(\ref{eq:corCexp}), whatever be the value of $\mu$. At a given time $t$, the correlation length $\xi _C(M,t) \propto t^{1/z'}$ in the time regime of interest, and $\la \sigma _M^2(t)\ra$ is a constant that does not depend on $k$. We therefore have, for $k \ll t^{1/z'}$,
\begin{equation}
G_M(k,t) \approx \la \sigma _M^2(t)\ra - K_C\left( 1 - \frac{k}{t^{1/z'}}\right)  
\label{eq:corGalpha}
\end{equation}
where $K_C$ is a proportionality constant. Comparing it with the expected scaling behavior, Eq.(\ref{eq:Gscale2}), we have the result $\alpha ' = 1/2$ independent of $\mu$.

For lag distances $k \gg t^{1/z'}$, we observe that $G_M(k,t)$ saturates to a value, say $G_M^{(s)}(t)$, after a lapse of time $t$. Correspondingly, we see that $C_M(k,t)$ is more or less saturated to a negative value, say -$C_M^{(s)}(t)$. The saturated values, estimated as averages of the last $M/8$ data points at various times, for $\mu = 1/2 \ \& \ 5$ and $M = 512$ are presented in Fig.(\ref{fig:GsCsSigma}a,b). Straight line fits to the data lead to the estimate $\beta ' \approx 1/4$. The corresponding estimates of $\la \sigma _M^2(t)\ra$ are also presented in these figures.  

Since in this time regime, $\la \sigma _M^2(t)\ra \sim t^{2\beta}$, we can write
\begin{equation}
\frac{1}{2}G_M^{(s)}(t) = K_{\sigma}t^{2\beta} + K_C t^{2\beta '} 
\label{eq:corGbeta}
\end{equation}
Which exponent will characterize the time-dependence of $G_M^{(s)}(t)$ is decided by the proportionality constants $K_{\sigma}$ and $K_C$. 

In the case $\mu = 1/2$, straight line fit to the values of $\la \sigma _M^2(t)\ra$ in Fig.(\ref{fig:GsCsSigma}a) give an estimate $\beta \approx \beta ' \approx 1/4$. Hence, $G_M^{(s)}(t) \propto t^{1/4}$ - independently supported by the Monte Carlo data as well.

On the other hand, for $\mu = 5$, $\beta \approx 1/2$ and $\beta ' \approx 1/4$ are the estimates obtained by fitting the data presented in Fig.(\ref{fig:GsCsSigma}b). The proportionality constants have the values, $K_{\sigma} \approx 2.5\times 10^{-4}$ and $K_C \approx 0.005$, respectively. From Eq.(\ref{eq:corGbeta}), it is clear that $G_M^{(s)}(t) \sim t^{2\beta}$ only for times $t \gg (K_C/K_{\sigma})^2 \approx 400$. However, data presented in Fig.(\ref{fig:SigmaScaleEW1}b) indicate that $\la \sigma _M^2(t)\ra$ saturates for $t \gg 400$. Hence, $G_M^{(s)}(t) \propto t^{1/4}$ - independently supported by the Monte Carlo data as well.
\begin{figure} [hbt] 
  \begin{center}
\includegraphics[width=3.5in,height=3.0in]{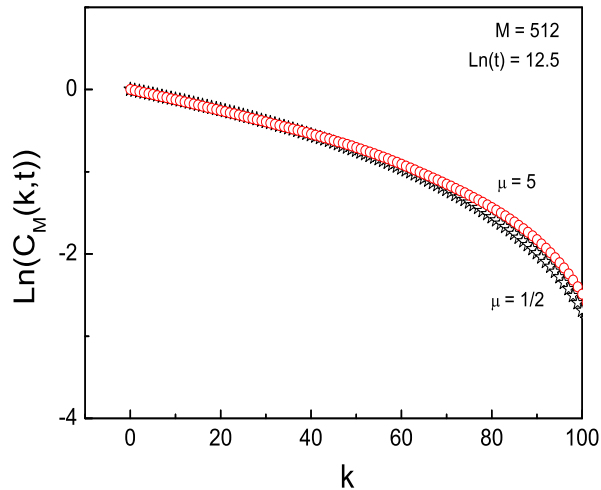}

    \caption{Semi-log plot of the Height-Fluctuation correlation function, $C_M(k,t)$, for $M = 512$ and 
     $\mu = 1/2 \ \& \ 5$. The data were obtained after a time $t = e^{12.5}$ Monte Carlo sweeps, and 
     averaged over $64$ independent runs. It is clear that $C_M(k,t)$ is an exponentially decaying 
     function over a reasonable range of lag distances ($k \in [0, \sim 48]$). 
    } 
    \label{fig:corC512}
  \end{center}
\end{figure}

\begin{figure} [hbt] 
  \begin{center}

    \subfigure[]{\includegraphics [width=0.48\textwidth] 
    {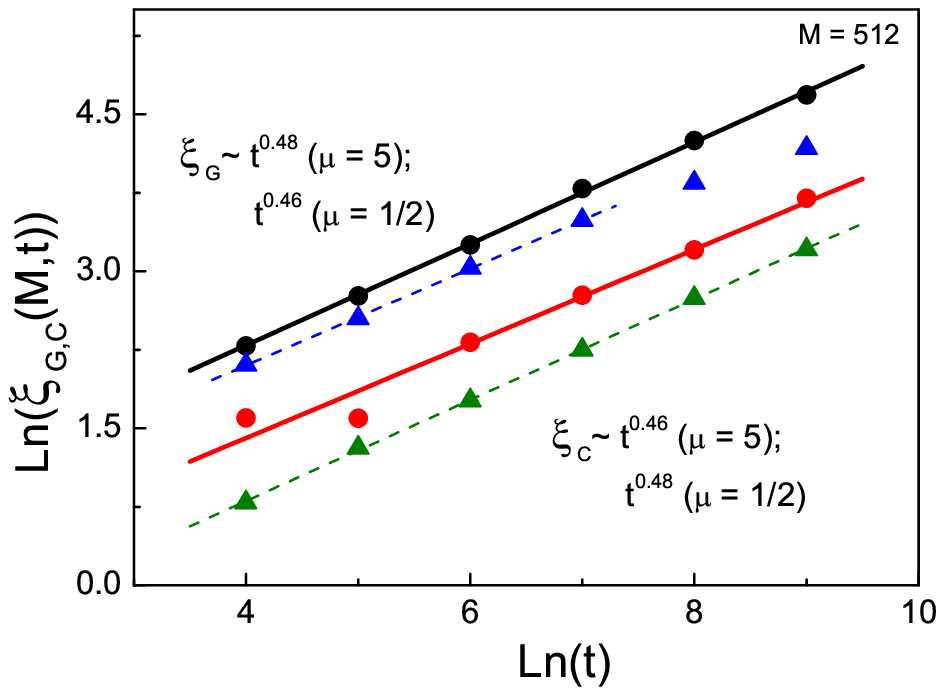}}
    \subfigure[]{\includegraphics [width=0.48\textwidth] 
    {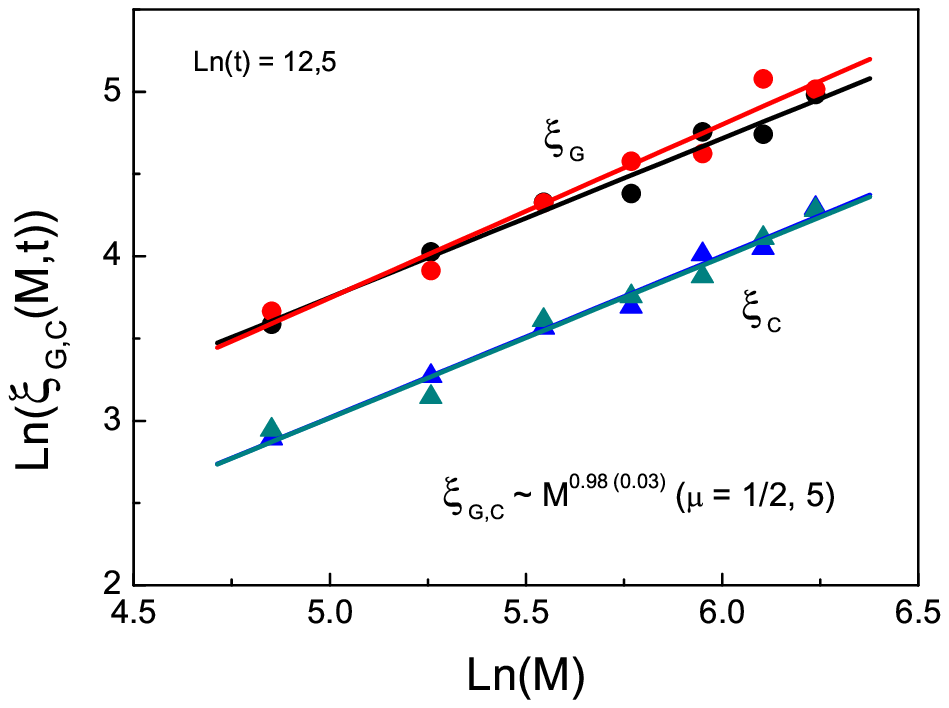}}       
    \caption{(a) Correlation lengths, $\xi _{G,C}(M,t)$, for a system of size $M = 512$ as a function of time. Filled circles correspond to $\xi _G(M,t)$, while filled triangles correspond to $\xi _C(M,t)$; upper ones are for $\mu = 5$ and the lower ones for $\mu = 1/2$. Within errors, $\xi _{G,C}(M,t) \sim t^{1/2}$. (b)Correlation lengths, $\xi _{G,C}(M,t)$, as a function of $M$ after a time $t = e^{12.5}$ Monte Carlo sweeps. Filled circles correspond to $\xi _G(M,t)$, while filled triangles correspond to $\xi _C(M,t)$; upper ones are for $\mu = 5$ and the lower ones for $\mu = 1/2$. Within errors, $\xi _{G,C}(M,t) \sim M$.    
    } 
    \label{fig:CorLenXiGC}
  \end{center}
\end{figure}

\begin{figure} [hbt] 
  \begin{center}            
\subfigure[]{\includegraphics[width=0.48\textwidth]{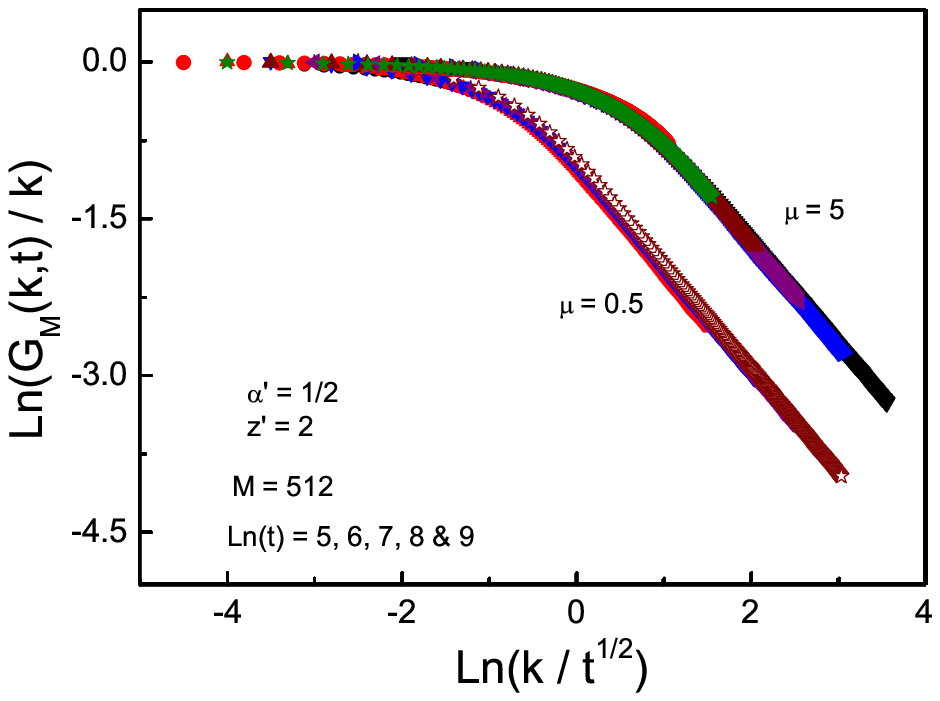}}
\subfigure[]{\includegraphics[width=0.48\textwidth]{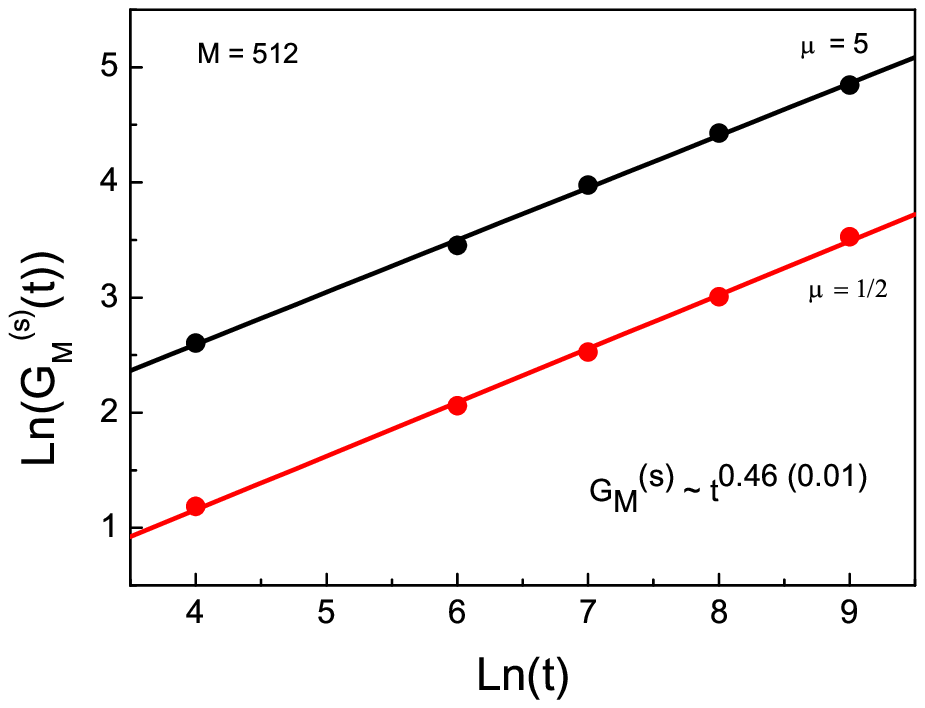}}

    \caption{(a) Scaling collapse of the HDCF, $G_M(k,t)$, estimated at times $ln(t) = 5, 6, 7, 8 \ \& \ 9$ for a system of size $M = 512$ and for $\mu = 1/2 \ \& \ 5$. It is evident that $\alpha ' = 1/2$ and $z' = 2$ independent of $\mu$.  (b) Saturation values of $G_M(k,t)$ at these times for $\mu = 1/2 \ \& \ 5$. Straight line fits give $\beta ' = 1/4$, within errors, independent of $\mu$.
    } 
    \label{fig:corG512collapse}
  \end{center}
\end{figure}

\begin{figure} [hbt] 
  \begin{center}            
\subfigure[]{\includegraphics[width=0.48\textwidth]{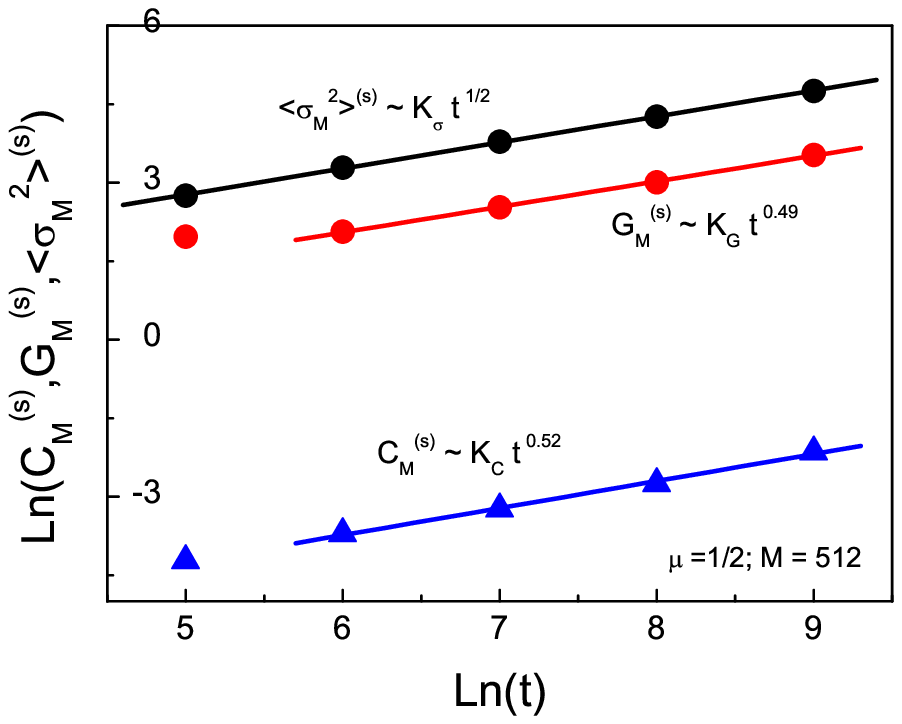}}
\subfigure[]{\includegraphics[width=0.48\textwidth]{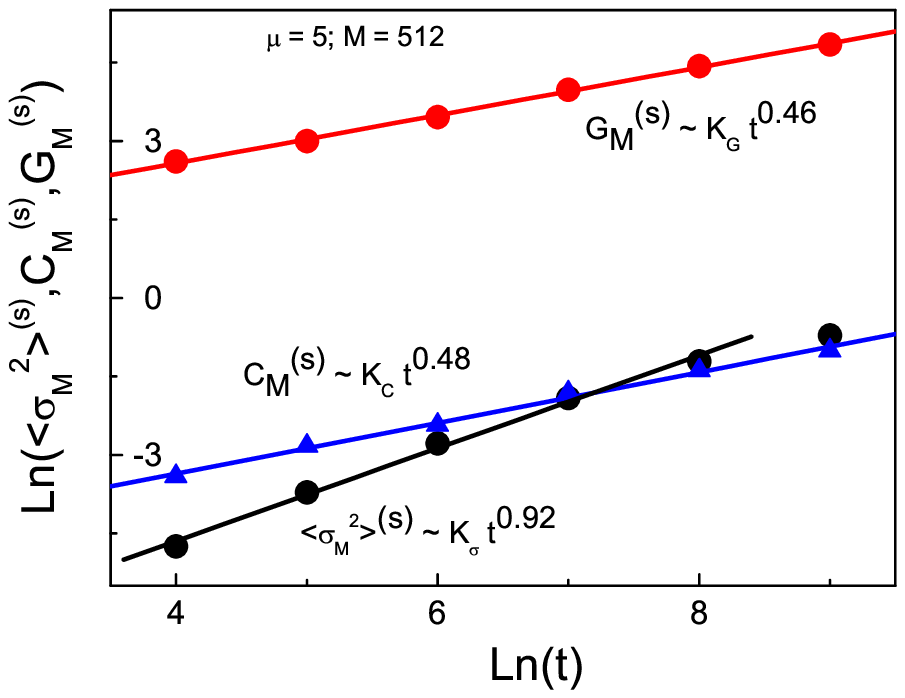}}

    \caption{Saturation values of the correlation functions, $G_M(k\gg t^{1/z},t)$, $C_M(k\gg t^{1/z},t)$, at various times $t (\ll M^z)$ and the corresponding mean squared widths, $\la \sigma _M^2(t)\ra$. (a) $\mu = 1/2$: The proportionality constants are $K_G \approx 0.415, \ K_C \approx 0.001$ and $K_{\sigma} \approx 1.33$. The exponents are $\beta \approx \beta ' \approx 1/4$. (b) $\mu = 5$: The proportionality constants are $K_G \approx 2.11, \ K_C \approx 0.005$ and $K_{\sigma} \approx 0.00025$. The exponents are $\beta \approx 1/2, \ \beta ' \approx 1/4$. 
    } 
    \label{fig:GsCsSigma}
  \end{center}
\end{figure}

\vspace*{0.5cm}
\subsection{Discussion.}

Metropolis dynamics of surface growth based on the SOS Hamiltonian, Eq.\ref{eq:energy}, has already been studied by Siegert and Plischke \cite{Plischke}. In their model, particle deposition is a random event that is extraneous to the energetics of the evolving surface, whereas diffusion of randomly deposited particles is what relaxes the surface towards an equilibrium energy state. Their model is motivated by the observation that relaxation of a surface, being grown by the molecular beam epitaxy process, is through the diffusion of particles on the surface. They have shown that the surface belongs to the EW class.

On the other hand, in the model we have presented here, the surface relaxes towards its equilibrium energy state not by the process of diffusion but by the process of particle deposition or evaporation. Since a change in energy is associated with an event of particle deposition or evaporation, we may say that this model treats a growing surface as a canonical object evolving towards its equilibrium state. Again, since equilibrium energy is the same as the equilibrium contour length (surface area, in higher dimensions) which, in turn, is equivalent to the equilibrium number of particles defining the surface, we may also say that this model treats a growing surface as a grand canonical object.

The motivation for this model is to provide a microscopic view of an equilibrium elastic membrane consisting of a number of folds of varying sizes for a given parameter $\mu$ (in thermal units). Therefore, diffusion cannot be the mechanism for surface relaxation in the SOS picture adopted here. The Monte Carlo study presented here provides a strong evidence that the parameter $\mu$ can tune the exponents, $\alpha$ and $\beta$, characterizing the scaling behavior of the surface-width.  

Normally, in small scale simulations ($M$ not very large), better estimates of these exponents are obtained from the scaling behavior of the height-difference correlation function (HDCF), $G_M(k,t)$, rather than from that of the surface-width. Jeong and Kim \cite{JK} have recently shown that this is true only for models such as the Restricted Solid-on-Solid (RSOS) model \cite{KK,KKA} that have no global constraints on the column heights. Using a restricted curvature model, they have demonstrated that the exponents of $G_M(k,t)$ are not the same as those of the surface width when the extremal height of the surface is suppressed.  

Eventhough no such global costraints are there in our model, the exponents of $G_M(k,t)$ are different from those of the surface-width for $\mu > 2$; the surface is self-constraining for higher values of $\mu$. The subtle point this study has brought out is the decisive influence of proportionality constants on the scaling behavior of $G_M(k,t)$.

For example, the time-dependence of the saturation value of $G_M(k,t)$ (typically, for $k > M/4$), as given by Eq.(\ref{eq:corGbeta}), depends crucially on the proportionality constants $K_{\sigma}$ and $K_C$. If $\beta > \beta '$, then for times $t \gg (K_C/K_{\sigma})^{1/2(\beta -\beta ')}$, $G_M^{(s)}(t) \propto t^{2\beta}$ provided the surface width has not yet saturated; else, $G_M^{(s)}(t) \propto t^{2\beta '}$. In our model, $\beta ' = 1/4$ independent of $\mu$, whereas $\beta = 1/2$ for $\mu > \sim 3.5$. 

Since the height-fluctuation correlation function (HFCF) $C_M(k,t)$ is an exponentially decaying function of the lag-distance $k$ (Eq.(\ref{eq:corCexp})), whatever be the value of $\mu$, the equality Eq.(\ref{eq:corGsigma}) for $G_M(k,t)$ can be rewritten as
\begin{equation}
G_M(k,t) \approx \la \sigma _M^2(t)\ra - \left( 1 - \frac{k}{\xi _C(M,t)}\right); 
                                                                     \quad k \ll \xi _C(M,t)
\label{eq:corGalpha1}
\end{equation}
In the growth regime ($t \ll M^z$), at a particular time $t$, the $k$-dependence of $G_M(k,t)$ is given by $G_M(k,t) \propto k$; this implies, by definition Eq.(\ref{eq:Gscale2}), that $\alpha ' = 1/2$, whatever be the value of $\mu$. In contrast, $\alpha$ increases with $\mu$ for $\mu > \sim 4$.

\section{Summary}

We have presented an SOS model of particle deposition and evaporation for describing the roughness of a $(1+1)$-dimensional equilibrium elastic membrane with folds of varying sizes. The stochastic evolution of an intial flat surface towards its equilibrium configuration is governed by the standard Metropolis criterion based on the {\it energy} (or equivalently, the {\it contour length}), given by Eq.(\ref{eq:energy}), associated with this surface. The only tunable parameter in this model is $\mu$, the dimensionless surface tension parameter.

For negative values of $\mu$, the surface becomes increasingly spiky and its width (root mean squared fluctuation of its heights), $\sigma \propto t$. But for $\mu = 0$, $\sigma \propto {\sqrt t}$. In either case, $\sigma$ does not saturate to a stationary value.

For $\mu > 0$, on the other hand, we observe an asymptotic saturation of $\sigma$ indicating that the surface evolves into an equilibrium state. In fact, our Monte Carlo data for $0 < \mu \leq 2$ show that the surface belongs to the Edwards-Wilkinson class characterized by the exponents, $\beta = 1/4$ and $\alpha = 1/2$. However, for $\mu = 3.5$, the growth exponent $\beta$ is different ($\beta = 1/2$) while the roughness exponent, $\alpha$, has the same Edwards-Wilkinson value. For $\mu > \sim 4$, we observe that $\alpha$ increases with $\mu$, while $\beta$ ($= 1/2$) remains constant. Our Monte Carlo study, thus, provides a strong evidence that the roughness exponent, $\alpha$, of the $(1+1)$-dimensional surface can be tuned by the surface tension parameter, $\mu$.

We have also shown that the exponents of the height-difference correlation function differ from those of the surface width for higher values of $\mu$ suggesting thereby that the surface could be self-constraining. We have presented a brief heuristic argument to support this observation. There is a definite need for a detailed analytical, and large scale numerical, study of this simple but rich model   





\end{document}